\documentclass[preprint,12pt]{elsarticle}

\newcounter{bla}

\journal{Computer Physics Communications}

\begin{document}

\begin{frontmatter}

\title{ RNGSSELIB: Program library for random number generation.
More generators, parallel streams of random numbers and Fortran compatibility.}

\author{L.Yu. Barash}
\ead{barash@itp.ac.ru}
\author{L.N. Shchur}
\address{Landau Institute for Theoretical Physics, 142432
Chernogolovka, Russia}
\address{Moscow Institute of Physics and Technology, 141700 Moscow, Russia}
\address{National Research University Higher School of Economics,
Moscow Institute of Electronics and Mathematics,
109028 Moscow, Russia}

\begin{abstract}
In this update, we present the new version of the random
number generator (RNG) library RNGSSELIB, which, in particular, contains
fast SSE realizations of a number of modern and most reliable
generators~\cite{RNGSSELIB1}. The new features are: i) Fortran compatibility and examples of
using the library in Fortran; ii) new modern and reliable generators;
iii) the abilities to jump ahead inside RNG sequence
and to initialize up to $10^{19}$ independent random number streams
with block splitting method.

\end{abstract}

\end{frontmatter}

{\bf NEW VERSION PROGRAM SUMMARY}

\begin{small}
\noindent
{\em Manuscript Title:}
RNGSSELIB: Program library for random number generation.
More generators, parallel streams of random numbers and Fortran compatibility.\\
{\em Authors:}  L.Yu. Barash, L.N. Shchur \\
{\em Program Title:}  RNGSSELIB   \\
{\em Journal Reference:}  Comput. Phys. Commun. 184 (2013) 2367  \\
{\em Catalogue identifier:}                                   \\
{\em Licensing provisions:}                                   \\
{\em Program summary URL:} \verb# http://cpc.cs.qub.ac.uk/summaries/AEIT_v2_0.html# \\
{\em Program obtainable from:} CPC Program Library, Queen's University, Belfast, N.~Ireland \\
{\em Licensing provisions:} Standard CPC license, \verb#http://cpc.cs.qub.ac.uk/licence/licence.html# \\
{\em No. of lines in distributed program, including test data, etc.:} 9299 \\
{\em No. of bytes in distributed program, including test data, etc.:} 1768030 \\
{\em Distribution format:} tar.gz \\
{\em Programming language:}  C, Fortran                       \\
{\em Computer:} PC, laptop, workstation, or server with Intel or AMD processor  \\
{\em Operating system:}  Unix, Windows                        \\
{\em RAM:} 4 Mbytes                                              \\
{\em Number of processors used:}                              \\
{\em Supplementary material:}                                 \\
{\em Keywords:}  Statistical methods, Monte Carlo,
  Random numbers, Pseudorandom numbers, Random number generation,
Streaming SIMD Extensions \\
{\em Classification:}  4.13 Statistical Methods   \\
{\em External routines/libraries:}                            \\
{\em Subprograms used:}                                       \\
{\em Catalogue identifier of previous version:} AEIT\_v1\_0   \\
{\em Journal reference of previous version:} Comput. Phys. Commun. 182 (2011) 1518  \\
{\em Does the new version supersede the previous version?:} Yes \\
{\em Nature of problem:} Any calculation requiring uniform pseudorandom
number generator, in particular, Monte Carlo calculations. Any calculation requiring 
parallel streams of uniform pseudorandom numbers.  \\
   \\
{\em Solution method:}
The library contains realization of the following modern and reliable generators:
\verb#MT19937#~\cite{MT19937}, \verb#MRG32K3A#~\cite{MRG32K3A}, \verb#LFSR113#~\cite{LFSR113},
\verb#GM19#, \verb#GM31#, \verb#GM61#~\cite{RNGSSELIB1}, and
\verb#GM29#, \verb#GM55#, \verb#GQ58.1#, \verb#GQ58.3#, \verb#GQ58.4#~\cite{EPL2011,Springer2012}.
The library contains both usual realizations and realizations based on SSE command set.
Usage of SSE commands allows to substantially improve performance of all generators.
Also, the updated library contains the abilities to jump ahead
inside RNG sequence and to initialize independent random number streams
with block splitting method for each of the RNGs.
\\
   \\
{\em Reasons for the new version:}
1. In order to implement Monte Carlo calculations, the implementation of
independent streams of random numbers is necessary. Such implementation
of initializing random number streams with block splitting method
is added to the new version for each of the RNGs. Jumping ahead
inside RNG sequence, which is necessary for the block splitting method,
was also added for each of the RNGs.
2. Users asked us to add Fortran compatibility to the library. Fortran
compatibility and the examples of using the library in Fortran for
each of the RNGs are included in this version.
3. During last few years, the method of random number generation based on
using the ensemble of transformations of two-dimensional torus,
was essentially improved~\cite{EPL2011,Springer2012}. Important properties,
such as high-dimensional equidistribution, were established for
the RNGs of this type. The proper choice of
parameters was determined, which resulted in the validity
of the high-dimensional equidistribution property,
and, correspondingly, the new high-quality RNGs
\verb#GM29#, \verb#GM55.4#, \verb#GQ58.1#,
\verb#GQ58.3#, \verb#GQ58.4# were proposed.
These RNGs are included now in the RNGSSELIB library.
\\
   \\
{\em Summary of revisions:}
1. We added Fortran compatibility and examples of
using the library in Fortran for each of the generators.
2. New modern and reliable generators \verb#GM29#, \verb#GM55.4#, \verb#GQ58.1#,
\verb#GQ58.3#, \verb#GQ58.4#, which were introduced in~\cite{EPL2011} were added to the library.
3. The abilities to jump ahead inside RNG sequence and
to initialize independent random number streams
with block splitting method are added for each of the RNGs.
\\
   \\
{\em Restrictions:} For SSE realizations of the generators,
Intel or AMD CPU supporting SSE2 command set is required.
In order to use the SSE realization for the \verb#lfsr113# generator,
CPU must support SSE4.1 command set. \\
   \\
{\em Unusual features:}\\
   \\
{\em Additional comments:}
The function call interface has been slightly modified compared
to the previous version in order
to support Fortran compatibility. For each of the generators,
RNGSSELIB supports the following functions, where \verb#rng# should
be replaced by the particular name of the RNG:

\begin{verbatim}
void rng_skipahead_(rng_state* state, unsigned long long offset);
void rng_init_(rng_state* state);
void rng_init_sequence_(rng_state* state,unsigned long long SequenceNumber);
unsigned int rng_generate_(rng_state* state);
float rng_generate_uniform_float_(rng_state* state);
unsigned int rng_sse_generate_(rng_sse_state* state);
void rng_get_sse_state_(rng_state* state,rng_sse_state* sse_state);
void rng_print_state_(rng_state* state);
void rng_print_sse_state_(rng_sse_state* state);
\end{verbatim}

There are a few peculiarities for some of the RNGs.
For example, the function\\
\verb#void mt19937_skipahead_(mt19937_state* state, unsigned long long#
\verb#a, unsigned b);#\\
skips ahead $N=a\cdot 2^b$ numbers, 
where $N<2^{512}$, and the function\\
\verb#void gm55_skipahead_(gm55_state* state, unsigned long long offset64,#\\ 
\verb#unsigned long long offset0);#\\
skips ahead $N=2^{64}\cdot\verb#offset64#+\verb#offset0#$ numbers.
The detailed function call interface can be found
in the header files of the \verb#include# directory.
The examples of using the library can be found in the
\verb#examples# directory.

\begin{table}
\caption{Initialization of pseudorandom streams for RNGs.}
\label{StreamsTable}
\begin{tabular}{|l|c|c|}
\hline
Function initializing sequence     & Number of sequences & Maximal length     \\
\hline
\verb#gm19_init_sequence_#             &    $1000$           & $6\cdot 10^6$      \\
\verb#gm29_init_short_sequence_#       &    $10^8$           & $8\cdot 10^7$      \\
\verb#gm29_init_medium_sequence_#      &    $10^6$           & $8\cdot 10^9$      \\
\verb#gm29_init_long_sequence_#        &    $10^4$           & $8\cdot 10^{11}$   \\
\verb#gm31_init_short_sequence_#       &    $10^9$           & $8\cdot 10^7$      \\
\verb#gm31_init_medium_sequence_#      &    $10^7$           & $8\cdot 10^9$      \\
\verb#gm31_init_long_sequence_#        &    $10^5$           & $8\cdot 10^{11}$   \\
\verb#gm55_init_short_sequence_#       &    $10^{18}$        & $10^{10}$          \\
\verb#gm55_init_long_sequence_#        &    $4\cdot 10^9$    & $10^{20}$          \\
\verb#gq58x1_init_short_sequence_#     &    $10^8$           & $8\cdot 10^7$      \\
\verb#gq58x1_init_medium_sequence_#    &    $10^6$           & $8\cdot 10^9$      \\
\verb#gq58x1_init_long_sequence_#      &    $10^4$           & $8\cdot 10^{11}$   \\
\verb#gq58x3_init_short_sequence_#     &    $2\cdot 10^8$    & $8\cdot 10^7$      \\
\verb#gq58x3_init_medium_sequence_#    &    $2\cdot 10^6$    & $8\cdot 10^9$      \\
\verb#gq58x3_init_long_sequence_#      &    $2\cdot 10^4$    & $8\cdot 10^{11}$   \\
\verb#gq58x4_init_short_sequence_#     &    $3\cdot 10^8$    & $8\cdot 10^7$      \\
\verb#gq58x4_init_medium_sequence_#    &    $3\cdot 10^6$    & $8\cdot 10^9$      \\
\verb#gq58x4_init_long_sequence_#      &    $3\cdot 10^4$    & $8\cdot 10^{11}$   \\
\verb#gm61_init_sequence_#             &  $1.8\cdot 10^{19}$ & $10^{10}$          \\
\verb#gm61_init_long_sequence_#        &    $4\cdot 10^9$    & $3\cdot 10^{25}$   \\
\verb#lfsr113_init_sequence_#          &  $3.8\cdot 10^{18}$ & $10^{10}$          \\
\verb#lfsr113_init_long_sequence_#     &    $4\cdot 10^9$    & $10^{24}$          \\
\verb#mrg32k3a_init_sequence_#         &       $10^{19}$     & $10^{38}$          \\
\verb#mt19937_init_sequence_#          &       $10^{19}$     & $10^{130}$         \\
\hline
\end{tabular}
\end{table}

Table 1 shows maximal number of sequences and maximal 
length of each sequence for each function
initializing pseudorandom stream.
The algorithms used to jump ahead in the RNG sequence and to
initialize parallel streams of pseudorandom numbers
are described in detail in~\cite{PrEng} and in~\cite{PRAND}.

This work was partially supported by Russian Foundation for Basic Research
projects No. 12-07-13121 and 13-07-00570
and by the Supercomputing Center of Lomonosov Moscow State University~\cite{Lomonosov}.
\\
   \\
{\em Running time:} Running time is of the order of $20$ sec
for generating $10^9$ pseudorandom numbers with a PC based
on Intel Core i7-940 CPU. Running time is analyzed in detail
in~\cite{RNGSSELIB1} and in~\cite{EPL2011}.
\\
   \\
\end{small}

\end{document}